\documentclass[twocolumn,showpacs,preprintnumbers,amssymb]{revtex4}

\usepackage{graphicx}
\usepackage{bm}

\begin{document}

\centerline{}
\title{Quasinormal modes of the near extremal
Schwarzschild-de Sitter black hole}

\author{Vitor Cardoso}
\email{vcardoso@fisica.ist.utl.pt}
\author{Jos\'e P. S. Lemos}
\email{lemos@kelvin.ist.utl.pt}
\affiliation{
Centro Multidisciplinar de Astrof\'{\i}sica - CENTRA, 
Departamento de F\'{\i}sica, Instituto Superior T\'ecnico,
Av. Rovisco Pais 1, 1049-001 Lisboa, Portugal
}%

\date{\today}

\begin{abstract}

We present an exact expression for the quasinormal modes of scalar,
electromagnetic and gravitational perturbations of a near extremal
Scwarzschild-de Sitter black hole and we show why a previous
approximation holds exactly in this near extremal regime.  In
particular, our results give the asymptotic behavior of the
quasinormal frequencies for highly damped modes, which has recently
attracted much attention due to the proposed identification of its
real part with the Barbero-Immirzi parameter.

\end{abstract}


\maketitle
\newpage
\section{Introduction}
The quasinormal modes of spacetimes containing black holes have proved
to be extremely important in several astrophysical aspects; for
instance, they allow us to ascertain whether or not the spacetime is
stable against deviations from equilibrium, and they give us invaluable 
information on what kind of signal one expects if one perturbs the
spacetime. In fact the quasinormal modes and their associated
frequencies are a signature of the spacetime, in that they depend only
on the conserved charges, such as the mass and electrical charge, making it
possible to identify the spacetime just by seeking its quasinormal
frequencies. This has motivated a wide effort to find the quasinormal
frequencies, and several numerical and analytical techniques have been
devised \cite{kokkotas}.  

It has been also realized that the
quasinormal modes are important in the context of the AdS/CFT
conjecture \cite{maldacena,horowitz0}.  According to it, the black hole
corresponds to a thermal state in the conformal field theory, and the
decay of the test field in the black hole spacetime corresponds to
the decay of the perturbed state in the CFT.  The dynamical timescale
for the return to thermal equilibrium is very hard to compute
directly, but can be done relatively easily using the AdS/CFT
correspondence. This has motivated a search for the quasinormal modes
in asymptotically anti-de Sitter black holes 
\cite{horowitz0,vitorjose,moss,all}.  

Very recently, the quasinormal modes have acquired a different
importance.  Following an observation by Hod \cite{hod}, it has been
proposed \cite{dreyerkunstattercorichi} that the Barbero-Immirzi
parameter \cite{immirzi}, a factor introduced by hand in order that
Loop Quantum Gravity reproduces correctly the black hole entropy, is
equal to the real part of the quasinormal frequencies with a large
imaginary part.  The identification came from what first seemed to be
a numerical coincidence, but which has been proved to be exact for
Schwarzschild black holes by Motl \cite{motl}, assuming the gauge
group of the theory to be $SO(3)$.

It is now important to see
whether the agreement works only for Schwarzschild black holes, or if
it continues to be true in different spacetimes, for example higher
dimensional Schwarzschild spacetime (where some of the lowest lying
scalar quasinormal frequencies have already been computed
\cite{vitoroscarjose}), or Schwarzschild-de Sitter or anti-de Sitter
spacetimes.  In this work we shall take a step further on carrying on
this program by computing exactly the quasinormal frequencies of the
near extremal Schwarzschild-de Sitter black hole, which is the
spacetime for which the black hole horizon and the cosmological
horizon are close to each other, in a manner to be defined latter.
For this spacetime, we find that it is possible to solve the field
equations exactly in terms of hypergeometric functions, and therefore
an exact analytical expression for the quasinormal frequencies of
scalar, electromagnetic and gravitational perturbations is also
possible. In particular this will give us the quasinormal frequencies
with very large imaginary part. We demonstrate why an approach by Moss
and Norman \cite{moss} based on fitting the potential to the
P\"oshl-Teller potential works well in the Schwarzschild-de Sitter
spacetime.

\section{Equations}
Our notation will follow that of \cite{brady} which we have found
convenient.  The metric of the Schwarzschild-de Sitter (SdS) spacetime
is given by
\begin{equation}
ds^2 = -f\, dt^2 + f^{-1}\, dr^2 + r^2 (d\theta^2 
+ \sin^2\theta\, d\phi^2), 
\label{2.1}
\end{equation}
where 
\begin{equation}
f = 1 - \frac{2M}{r} - \frac{r^2}{a^2},
\label{2.2}
\end{equation}
with $M$ denoting the black-hole mass, and $a^2$ is given in terms of the
cosmological constant $\Lambda$ by $a^2 = 3/\Lambda$.  The spacetime
possesses two horizons: the black-hole horizon is at $r=r_b$ and the
cosmological horizon is at $r = r_c$, where $r_c > r_b$. The function
$f$ has zeroes at $r_b$, $r_c$, and $r_0 = -(r_b + r_c)$. In terms of
these quantities, $f$ can be expressed as
\begin{equation}
f = \frac{1}{a^2 r}\, (r-r_b)(r_c-r)(r-r_0).
\label{2.3}
\end{equation}
It is useful to regard $r_b$ and $r_c$ as the two fundamental
parameters of the SdS spacetime, and to express $M$ and $a^2$ as
functions of these variables. The appropriate relations are
\begin{equation}
a^2 = {r_b}^2 + r_b r_c + {r_c}^2
\label{2.4}
\end{equation}
and
\begin{equation}
2M a^2 = r_b r_c (r_b + r_c).
\label{2.5}
\end{equation}
We also introduce the surface gravity $\kappa_b$ associated with the
black hole horizon $r = r_b$, as defined by the relation 
$\kappa_b = \frac{1}{2} 
df/dr_{r=r_b}$. Explicitly,  we have
\begin{equation}
\kappa_b = \frac{ (r_c-r_b)(r_b-r_0) }{ 2a^2 r_b }.
\label{surface}
\end{equation}
After a Fourier decomposition in frequencies and a multipole expansion, 
the scalar, electromagnetic and gravitational perturbations all
obey a wave equation of the form \cite{vitorjose,all}
\begin{equation}
\frac{\partial^{2} \phi(\omega,r)}{\partial r_*^{2}} +
\left\lbrack\omega^2-V(r)\right\rbrack
\phi(\omega,r)=0 \,,
\label{waveequation}
\end{equation}
where the tortoise coordinate is given by 
\begin{equation}
r_* \equiv \int f^{-1}\, dr\,, 
\label{tortoise}
\end{equation}
and the potential $V$
depends on the kind of field under consideration. 
Explicitly, for scalar
perturbations 
\begin{equation}
V_{\rm s}=f\left[\frac{l(l+1)}{r^2}+\frac{2M}{r^3}-\frac{2}{a^2}\right] \,,
\label{potentialscalar}
\end{equation}
while for electromagnetic perturbations
\begin{equation}
V_{\rm el}=f\left[\frac{l(l+1)}{r^2}\right] \,.
\label{potentialelectromagnetic}
\end{equation}
The gravitational perturbations decompose into two sets \cite{vitorjose}, 
the odd and the even parity one. We find however that for this spacetime,
they both yield the same quasinormal frequencies, so it is enough
to consider one of them, the odd parity ones say, for which the potential is 
\cite{vitorjose}
\begin{equation}
V_{\rm grav}=f\left[\frac{l(l+1)}{r^2}-\frac{6M}{r^3}\right] \,.
\label{potentialgravitational}
\end{equation}
In all cases, we denote by $l$ the angular quantum number,
that gives the multipolarity of the field.
\subsection{The Near Extremal SdS Black Hole}
Let us now specialize to the near extremal SdS black hole, which is
defined as the spacetime for which the cosmological horizon $r_c$
is very close (in the $r$ coordinate) to the black hole horizon $r_b$, i.e.
$\frac{r_c-r_b}{r_b}<<1$.
For this spacetime one can make the following approximations
\begin{equation}
r_0 \sim -2r_{b}^2\,\,;\,a^2\sim 3r_{b}^2;\,\,
M \sim \frac{r_b}{3}\,\,;\,\kappa_b \sim \frac{r_c-r_b}{2r_{b}^2}\,.
\label{approximation1}
\end{equation}
Furthermore, and this is the key point, since $r$ is constrained to
vary between $r_b$ and $r_c$, we get $r-r_0 \sim r_b -r_0 \sim 3r_0$
and thus 
\begin{equation}
f \sim \frac{(r-r_b)(r_c-r)}{r_{b}^2}\,.
\label{approximation2}
\end{equation}
In this limit, one can invert the relation $r_*(r)$ 
of (\ref{tortoise}) to get
\begin{equation}
r= \frac{r_c e^{2\kappa_b r_*}+r_b}{1+e^{2\kappa_b r_*}}\,.
\label{rtortoise}
\end{equation}
Substituting this on the expression (\ref{approximation2})
for $f$ we find 
\begin{equation}
f = \frac{(r_c-r_b)^2}{4r_{b}^2\cosh{(\kappa_b r_*)}^2}\,.
\label{approximation3}
\end{equation}
As such, and taking into account the functional form of the potentials
(\ref{potentialscalar})-(\ref{potentialgravitational}) we see that for
the near extremal SdS black hole the wave equation (\ref{waveequation}) is 
of the form
\begin{equation}
\frac{\partial^{2} \phi(\omega,r)}{\partial r_*^{2}} +
\left\lbrack\omega^2-\frac{V_0}{\cosh{(\kappa_b r_*)}^2}\right\rbrack
\phi(\omega,r)=0 \,,
\label{waveequation2}
\end{equation}
with
\begin{equation}
V_0=\left\{ \begin{array}{ll}
            \kappa_{b}^2 l(l+1)\,,   & {\rm \,scalar\, and\, electromagnetic}\\
                                  &  {\rm perturbations}. \\ 
            \kappa_{b}^2 (l+2)(l-1)\,,   &{\rm \, gravitational} \\
                                     &  {\rm perturbations}\,.
\end{array}\right.
\label{V0}
\end{equation}
The potential in (\ref{waveequation2}) is the well known 
P\"oshl-Teller potential \cite{teller}. The solutions to 
(\ref{waveequation2})
were studied and they are of the hypergeometric type, 
(for details see Ferrari and Mashhoon \cite{ferrari}).
It should be solved under appropriate boundary conditions:
\begin{eqnarray}
\phi \sim e^{-i\omega r_*} \,,r_* \rightarrow -\infty \\
\phi \sim e^{i\omega r_*}  \,,r_* \rightarrow \infty. 
\label{behavior1}
\end{eqnarray}
These boundary conditions impose a non-trivial condition on
$\omega$ \cite{ferrari}, and those that satisfy both simultaneously
are called quasinormal frequencies. For the P\"oshl-Teller potential
one can show \cite{ferrari} that they are given by
\begin{equation}
\omega=\kappa_{b} \left [ -(n+\frac{1}{2})i+
\sqrt{\frac{V_0}{\kappa_b^2}-\frac{1}{4}} \right ]\,, n=0,1,...\,.
\label{solution}
\end{equation}
Thus, with (\ref{V0}) one has 
\begin{equation}
\frac{\omega}{\kappa_b}=-(n+\frac{1}{2})i+
\sqrt{l(l+1)-\frac{1}{4}}\,,n=0,1,...\,.
\label{finalsclarelectr}
\end{equation}
for scalar and electromagnetic perturbations.
And
\begin{equation}
\frac{\omega}{\kappa_b}=-(n+\frac{1}{2})i+\sqrt{(l+2)(l-1)-
\frac{1}{4}}\,,n=0,1,...\,.
\label{finalgrav}
\end{equation}
for gravitational perturbations.
Our analysis shows that
Eqs. (\ref{finalsclarelectr})-(\ref{finalgrav}) are correct up to
terms of order $O(r_c-r_b)$ or higher.  Moss and Norman \cite{moss} have
studied the quasinormal frequencies in the SdS geometry numerically
and also analytically, by fitting the potential to a P\"oshl-Teller
potential. Their analytical results (see their Figs 1 and 2) were in
excellent agreement with their numerical results, and this agreement
was even more remarkable for near extremal black holes and for high
values of the angular quantum number $l$.  We can now understand why:
for near extremal black holes the true potential is indeed given by
the P\"oshl-Teller potential!  Furthermore for near extremal SdS black
holes and for high $l$ our formula (\ref{finalgrav}) is approximately
equal to formula (19) of Moss and Norman \cite{moss}. With their
analytical method of fitting the potential one can never be sure if
the results obtained will continue to be good as one increases the
mode number $n$. But we have now proved that if one is in the near
extremal SdS black hole, the P\"oshl-Teller is the true potential, and
so Eq. (\ref{finalsclarelectr})-(\ref{finalgrav}) is exact.  For
example, Moss and Norman obtain numerically, and for gravitational
perturbations with $l=2$ of nearly extreme SdS black holes, the result
\begin{equation}
\frac{\omega_{\rm num}}{\kappa_b}=1.93648-i(n+\frac{1}{2})\,,
\label{mossnumel2}
\end{equation}
and we obtain, from (\ref{finalgrav})
\begin{equation}
\frac{\omega}{\kappa_b}=1.936492-i(n+\frac{1}{2})\,.
\label{meuexactl2}
\end{equation}
For $l=3$ Moss and Norman \cite{moss} obtain
\begin{equation}
\frac{\omega_{\rm num}}{\kappa_b}=3.12249-i(n+\frac{1}{2})\,,
\label{mossnumel3}
\end{equation}
and we obtain, from (\ref{finalgrav})
\begin{equation}
\frac{\omega}{\kappa_b}=3.122499-i(n+\frac{1}{2})\,.
\label{meuexactl3}
\end{equation}
So this remarkable agreement allows us to be sure that 
(\ref{finalsclarelectr})-(\ref{finalgrav}) are indeed correct.
\section{conclusions}
We have found an analytical expression for the quasinormal modes and
frequencies of a nearly extreme schwarzschild-de Sitter black
hole. This expression, Eqs. (\ref{finalsclarelectr})-(\ref{finalgrav})
are correct up to terms of order $O(r_c-r_b)$ or higher for all
$n$. This means that we can be confident that for high overtones,
i.e. large $n$, our expression is still valid.  One can see that the
real part of the quasinormal frequency does not depend on the integer
$n$ labelling the mode. Therefore, frequencies with a large imaginary
part still have a real part given by $\sqrt{l(l+1)-\frac{1}{4}}$ for
scalar and electromagnetic perturbations and by
$\sqrt{(l+2)(l-1)-\frac{1}{4}}$ for gravitational perturbations.  Can
one explain an highly damped quasinormal frequency with an
$l$-dependent real part in light of the recent conjectures
\cite{dreyerkunstattercorichi} relating it to the Barbero-Immirzi
parameter? We think it is too early to answer this, and much more work is
still necessary, specially in higher dimensional spacetimes, and on 
Anti-de Sitter spacetimes \cite{vitorromanjose}, where the AdS/CFT conjecture may 
have a word to say about this.

\vskip 2mm

\section*{Acknowledgements}
It is a pleasure to acknowledge stimulating conversations with
\'Oscar Dias and Lubos Motl. We are specially grateful to Ian Moss and
James Norman, who kindly provided their numerical results.  This work
was partially funded by Funda\c c\~ao para a Ci\^encia e Tecnologia
(FCT) -- Portugal through project PESO/PRO/2000/4014. V.C.  also
acknowledges finantial support from FCT through PRAXIS XXI programme.
J. P. S. L. thanks Observat\'orio Nacional do Rio de Janeiro for
hospitality.


\end{document}